\documentstyle[epsf,bibnorm]{lamuphys}
\makeatletter
\let\chapter\hid@chapter
\makeatother
\begin{document}
\pagenumbering{arabic}

\title{Coulomb breakup of drip-liners}

\author{Alberto Mengoni\inst{1,2}}

\institute{
National Institute for New Technology, Energy
and Environment (ENEA) \\ 
Applied Physics Division, Via Don Fiammelli, 2 - 40129 Bologna (Italy) 
\and
Institute for Physical and Chemical Research (RIKEN) \\
Radiation Laboratory, 2-1 Hirosawa, Wako, Saitama 351-0198 (Japan)
}


\maketitle

\begin{abstract}
We show how to derive nuclear structure information 
on loosely bound nuclei far from 
the $\beta$ stability line from their Coulomb dissociation 
into the continuum. 
The $^{19}{\rm C} \rightarrow {}^{18}{\rm C} + n$
process is taken as an example and it is shown that
basic properties such as ground-state spin
and parity and the neutron binding energy
of the halo nucleus $^{19}{\rm C}$ 
can be derived from a comparison of theoretical
calculations to experimental data. 
\end{abstract}

\section{Introduction}
The structure of light neutron-rich nuclei close
to the neutron drip-line exhibits some peculiar
properties which have been the subject of numerous
investigations in the most recent years.
For example, a low-density extended valence
neutron distribution decoupled from an
ordinary nuclear core has been found in nuclei
such as ${}^{11}{\rm Be}$ and ${}^{11}{\rm Li}$.
This property, the neutron halo structure,
is strongly related to the extremely
loose nature of the valence neutrons
and to their strong spectroscopic parentage
with the $2s_{1/2}$ single-particle orbit.

Coulomb breakup experiments have been
used to determine the halo structure
properties of neutron-rich light systems.
The Coulomb breakup (or Coulomb dissociation)
of a ${}^{A}{\rm X}$ nucleus 
into the ${}^{A-1}{\rm X}+n$ channel
is realized by using a high energy 
(typically radioactive ion) beam
on a high-$Z$ target. The virtual photons
generated by the Coulomb field of the
target are absorbed by the incident nucleus
causing an electromagnetic transition
from a bound state $|^{A}{\rm X}>$
into the $|^{A-1}{\rm X} + n>$
continuum. Here, $|^{A-1}{\rm X} + n>$
denotes a positive-energy 
(or scattering) state of the neutron
in the core of the halo nucleus.
As in the case of bound nuclear states
of stable or near-stable nuclei,
electromagnetic transition matrix
elements carry the most important
information on the nuclear wave functions
and hence on the structure.
It follows that basic structure
information are carried by the
Coulomb dissociation cross section
and strength distributions.

Here we will analyze the case of the
Coulomb breakup of a halo candidate,
${}^{19}{\rm C}$. This nucleus has been
the subject of recent experimental
and theoretical investigations because
its structure presents some
intriguing features.
So far, nuclear breakup experiments 
performed at MSU~\cite{BAZI95,BAZI98,BAUM98} 
and GSI~\cite{BAUM98} have failed to
provide a definitive answer even
for basic properties such as
the ground-state spin and parity
and the neutron separation energy.
The most recent experiment performed
in RIKEN~\cite{Nax98} seems to have solved these
two questions but many more on the
structure of this exotic system
seem to have arisen. We will
explore these here, and we will
show what is the theoretical
ground for their definition.

\section{Neutron rich Carbon isotopes}
\begin{figure}[t]
\centerline{
\epsfysize=8.0cm, 
\epsfbox{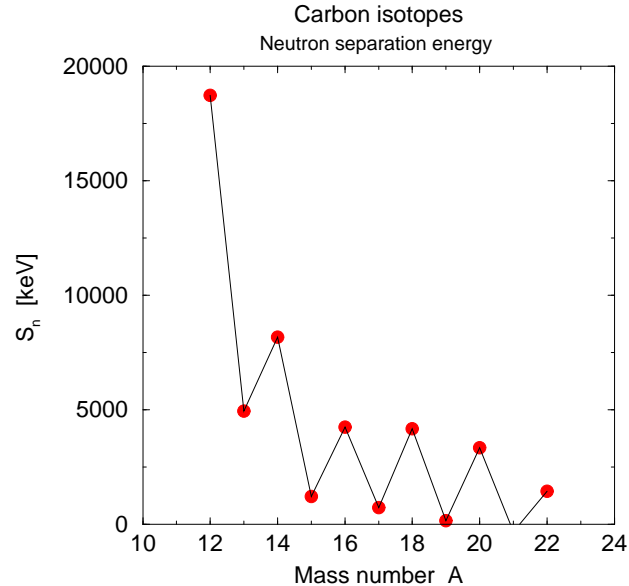}  
}
\caption[ ]{One-neutron separation energy for
Carbon isotopes with $N \geq 6$. $^{21}{\rm C}$ is
unbound and therefore $^{19}{\rm C}$ is the heaviest
even-odd bound isotope of the chain.}
\hfill
\vfill
\label{CBn}
\end{figure}

As previously mentioned, the halo
structure of light neutron-rich
nuclei is related to their loosely
bound nature. 
In Figure~\ref{CBn}, the one-neutron binding energy
of the $N \geq 6$ Carbon isotopes is shown as a function
of the mass number. Naturally, the binding energy
decreases as the neutron drip-line is approached.
The last even-odd bound isotope is $^{19}{\rm C}$
with a neutron binding energy of
$S_{n} = 160 \pm 110 $ KeV~\cite{AuWa95}.
A more reliable estimate of his quantity,
based on earlier mass measurements, is 
$S_{n} = 240 \pm 100 $ KeV~\cite{VIEI86,GILL87,WOUT88,ORR91}

\begin{figure}[t]
\centerline{
\epsfxsize=10cm 
\epsfysize=4cm 
\epsfbox{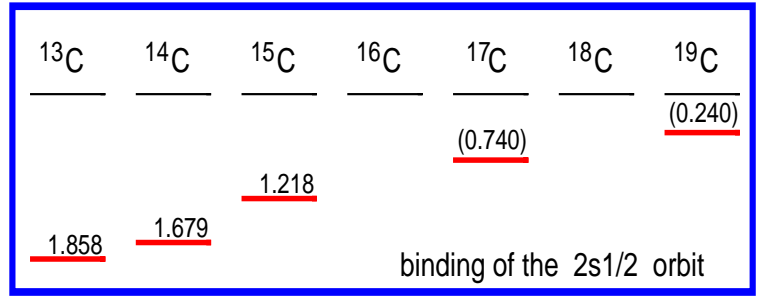}
}
\caption[ ]{Binding energy of the $2s_{1/2}$ single-particle
orbital in the Carbon isotopic chain. The value for 
$^{14}{\rm C}$ is the centroid of the two 
$J^{\pi} = 0^{-}, 1^{-}$ states. The values shown for
$^{17}{\rm C}$ and $^{19}{\rm C}$ are just the one-neutron 
separation energy of their ground-state.} 
\hfill
\vfill
\label{CBn_2s0.5}
\end{figure}

In the shell model ordering, $^{19}_{13}{\rm C}_{6}$
has five neutrons in the $sd$ neutron shell. In the
extreme single-particle picture they should occupy
the $1d_{5/2}$ orbital. However, mixing with the
$2s_{1/2}$ orbital is expected from the systematics
of this single-particle state in the Carbon isotopes, 
as shown in the Figure~\ref{CBn_2s0.5}. In the figure,
the binding energy of the $2s_{1/2}$ orbit in
the isotopic chain under consideration here is
shown. It has to be reminded that, already
for $^{13}{\rm C}$ this orbit is located
{\rm below} the $1d_{5/2}$. In addition,
the binding itself is becoming looser and
looser as the neutron number increases.
This trend supports the argument which would 
favor a halo structure for $^{19}{\rm C}$.

In the extreme single particle picture,
the structure of the $^{19}{\rm C}$
ground-state is given by
\begin{equation}
\Psi_{b}(J^{\pi}) = |^{18}{\rm C}(0^{+}) \otimes (nlj);j^{(-)^{l}}>
\end{equation}
where $nlj$ are single-particle orbital 
quantum numbers. Because of the strong effect
of the orbital angular momentum on the
amplitude of the tail of the radial 
wave function in the asymptotic region,
we can expect a Coulomb dissociation
cross section and $B({\rm E}\lambda)$
strength distribution completely
different for the allowed configurations
of the $^{19}{\rm C}$ ground state.
We can therefore expect to derive important
information on the structure of this
extremely neutron-rich nucleus
from the analysis of its Coulomb breakup.

\section{Coulomb dissociation}
The Coulomb dissociation cross section is 
related to the Coulomb field of the
target nucleus by~\cite{BeBa88}
\begin{equation}
\frac{d\sigma_{CD}}{dE_x} =                                                   
\sum_{\lambda} \ 
\int 2\pi b db \frac{N_{E\lambda}(E_{x},b)}{E_x}   
\sigma_{\gamma,x}^{E\lambda}(E_x)                                            
\end{equation}
where $N_{{\rm E}\lambda}(E_x,b)$ is the virtual 
photons number at impact parameter $b$ and
excitation energy $E_x$ 
(defined as the sum of the neutron-residual nucleus
relative kinetic energy plus the neutron binding energy)
and $\sigma_{\gamma,n}^{{\rm E}\lambda}(E_x)$
the photo-disintegration cross section. We will
specialize here to dipole radiation, i.e. $\lambda=1$.
The photo-disintegration cross section is given by
\begin{equation}
\sigma_{\gamma,n}^{\rm E1}(E_x) =
\frac{16 \pi}{9} \frac{E_x}{\hbar c} 
\frac{\mu k_{n}}{\hbar^2} \ \bar{e}^2 \
\frac{ 2J_{c} + 1}{2J_{b}+1} \ 
\vert Q^{({\rm E1})}_{b \rightarrow c} \vert^2
\end{equation}
where $\mu$ is the reduced mass of the system,
$k_{n}$ is the wave number of the
neutron-residual nucleus relative motion 
in the continuum,
$\bar{e}$ the E1 effective charge,
$J_{b}$ is the total angular momentum
of the bound state and $J_{c}$ the spin 
of the residual nucleus in the continuum. 

The photo-disintegration cross section can
be promptly related to the neutron capture
cross section by detailed balance
\begin{equation}
\sigma_{\gamma,n} = 
\frac{k_{n}^2}{k_{\gamma}^2} \
\frac{ 2J_{c} + 1}{2J_{b}+1} \ 
\sigma_{n,\gamma}.
\end{equation}
The essential ingredients for the calculation of the
Coulomb dissociation cross section 
are the matrix elements
\begin{equation}
{{\it Q^{({\rm E1})}_{b \rightarrow c}}} =
< \Psi _{c} \vert \hat{T}^{\rm E1} \vert \Psi _{b} >
\end{equation}
where $\Psi_{b}$ is the bound-state
wave function and $\Psi_{c}$ the wave function
for the neutron in the continuum. 
These matrix elements can be easily evaluated
for bound states with configurations of type
\mbox{$|{}^{A}\mbox{X}(J_{c}^{\pi})
\otimes (nlj); J_{b}>$}.
In this case, they can be decomposed into
the products of three factors
\begin{equation}
Q^{({\rm E1})}_{b \rightarrow c} \equiv
\sqrt{S_b} \ A_{b,c} \ {\cal I}_{b,c} 
\end{equation}
where $S_b$ is the spectroscopic factor of the bound state,
$A_{b,c}$ is a factor containing only angular momentum and
spin coupling coefficients and ${\cal I}_{b,c}$
the radial overlap. ${\cal I}_{b,c}$
can be evaluated using some potential model for the
calculation of the radial wave functions $u_{b}(r)$
and $w_{c}(r)$. 

\section{$^{19}{\rm C} \rightarrow {}^{18}{\rm C} + n$}
A first set of calculations of the Coulomb
dissociation cross section for the
$^{19}{\rm C} \rightarrow {}^{18}{\rm C} + n$
process can be performed using a simple Woods-Saxon
potential. The single-particle orbits
and their relative radial wave functions
can be easily calculated. The geometrical
parameters of the Woods-Saxon potential
well used for this calculation were $r_{0}=1.236 $ fm
(radius), $d=0.62$ fm (diffuseness). In
addition, a spin-orbit strength
$V_{so}=7.0$ MeV was used.
With these parameters, the
depth of the well was adjusted in order
to reproduce a neutron binding
energy of 0.240 MeV. As expected,
the depths corresponding to the
$2s_{1/2}$ and to the $1d_{5/2}$
were almost identical
($V_{0}=40.14$ MeV and $V_{0}=40.31$,
respectively). The almost-degeneracy of
these two orbits was expected.

\begin{figure}[t]
\centerline{
\epsfxsize=8.0cm 
\epsfysize=6.0cm 
\epsfbox{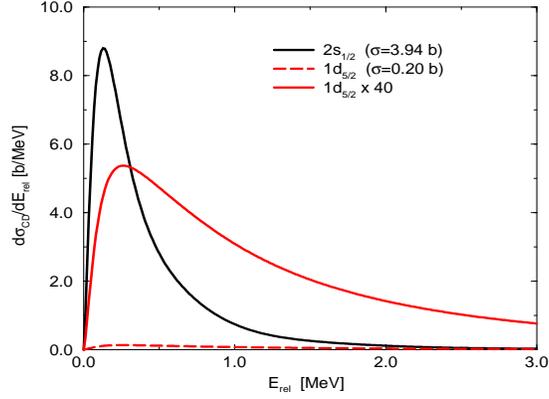}
}
\caption[ ]{Coulomb dissociation cross section of $^{19}{\rm C}$
on a Pb target at 67$A$ MeV. Only the dominant
${\rm E}1$ contribution is included. See text for details
on the other parameters used for the calculation.} 
\hfill
\vfill
\label{CD_sp}
\end{figure}

The result of the Coulomb dissociation
cross section are shown in Figure~\ref{CD_sp}.
It can be seen that the absolute magnitude as
well as the spectral shapes of the two
different configurations are completely
different. From these calculation it is
apparent that a measurement of the Coulomb
breakup of $^{19}{\rm C}$ would allow
to discriminate between a $J^{\pi}=1/2^{+}$
and $J^{\pi}=5/2^{+}$ ground state.

However, this simple approximation cannot
be considered sufficient for an accurate
description of the $^{19}{\rm C}$ structure.
We have then calculated the ground-state
structure using the shell-model. The
model space was a $\pi(p)\nu(sd)$
space with two different sets of
interactions, WBP and WBT~\cite{WaBr92}.

\begin{table}[h]                                                              
\caption{Shell model configurations of the $^{19}{\rm C}$ 
ground state, calculated using OXBASH~\protect\cite{OXBASH} with
WBP interaction. Only the configurations relevant
for the present study are shown. The full set of
results can be obtained upon request to the author.}
\begin{center} 
\begin{tabular}{| c | c  c  c |}
\hline 
$\epsilon_{1/2}-\epsilon_{5/2}$ & \ $J^{\pi}$ & 
 $S_{b}$ & configuration \\
   MeV    &           &               &         \\     
\hline                                                                        
--2.22 & $\frac{1}{2}^{+}$ & 0.58 &
$\vert ^{18}$C(0$^{+}$)$\otimes 2s_{1/2} \rangle$ \\ 
& & 0.47 &
$\vert ^{18}$C(2$^{+}$)$\otimes 1d_{5/2} \rangle$ \\               
\hline                                                                        
--0.38 & $\frac{5}{2}^{+}$ & 0.32 &
$\vert ^{18}$C(0$^{+}$)$\otimes 1d_{5/2} \rangle$ \\          
& & 1.16 &
$\vert ^{18}$C(2$^{+}$)$\otimes 1d_{5/2} \rangle$ \\               
& & 0.024 & 
$\vert ^{18}$C(2$^{+}$)$\otimes 2s_{1/2} \rangle$ \\
\hline                                                                        
\end{tabular}                                                                 
\end{center}
\label{onlytab}
\end{table}                                                                 
The full results of these calculations
have been reported elsewhere~\cite{TagTh98} 
and are available upon request.
Here we show only the results concerning
the most important configurations obtained
using the WBP interaction. In all cases,
a $^{14}_{8}{\rm C}_{6}$ core was assumed
(5 neutrons in the $sd$ shell).

In the table, two different set of calculations are shown
for two values of the $2s_{1/2} - 1d_{5/2}$
single particle energy difference. The first
one is the original value used in conjunction with
the WBP interaction while the second value
correspond more closely to our situation
in which the $s$ and $d$ orbitals are almost degenerate.
In the second case, the $^{19}{\rm C}$ 
ground state turns out to be a $J^{\pi}=5/2^{+}$
while with the $2s_{1/2}$ orbit 2.22 MeV below the 
$1d_{5/2}$ we obtained $J^{\pi}=1/2^{+}$. 
This result can be understood because by lowering
the $2s_{1/2}$ orbit one favors the 
$\vert ^{18}$C(0$^{+}$)$\otimes 2s_{1/2} \rangle$
configuration.

\begin{figure}[t]
\centerline{
\epsfxsize=10.0cm 
\epsfysize=10.0cm 
\epsfbox{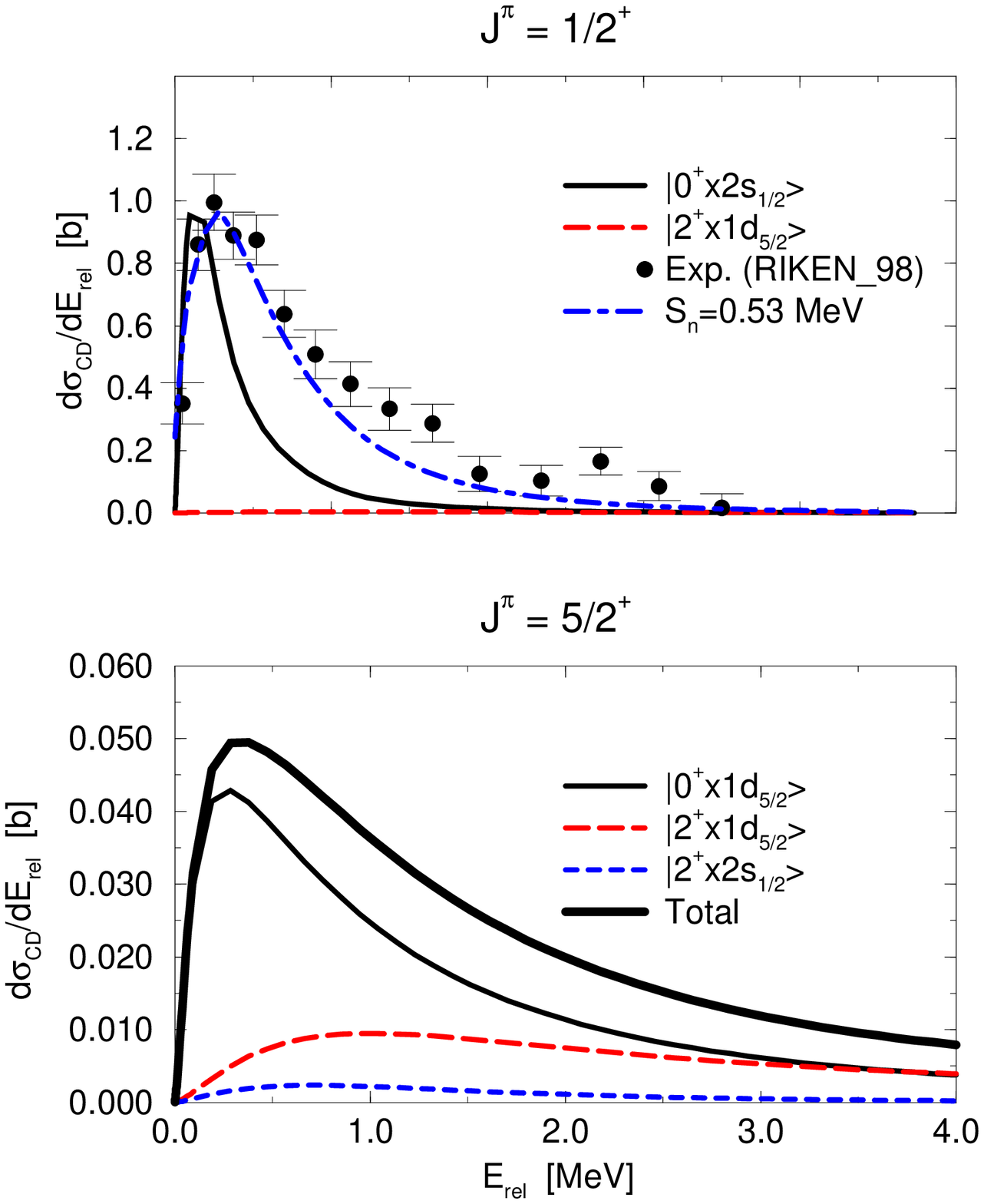}
}
\caption[ ]{Coulomb dissociation cross section of $^{19}{\rm C}$
on a Pb target at 67$A$ MeV. Only the dominant
${\rm E}1$ contribution is included. The experimental
data are from~\cite{Nax98}. See text for details
on the other parameters used for the calculation.} 
\hfill
\vfill
\label{CD_cm}
\end{figure}

The results of the Coulomb dissociation cross section
calculation are shown in Figure~\ref{CD_cm}.
There, a comparison with a recent experiment~\cite{Nax98}
performed at RIKEN RIPS facility is shown.
All the configurations resulting from
shell model calculations (see Table~\ref{onlytab})
were included in this calculation. 
From the comparison of the upper part with the lower part
of the figure it is apparent that a $J^{\pi}=5/2^{+}$
can be excluded for the $^{19}{\rm C}$ 
ground state. In fact, neither the absolute value
nor the spectral shape of the experimental cross section
could be reproduced with this assumption. Indeed,
even allowing for larger spectroscopic amplitudes
of the configurations included (up to full strengths),
the experimental cross section cannot be reproduced
with a $J^{\pi}=5/2^{+}$ assumption.
The integrated cross section for the calculations
obtained with $J^{\pi}=5/2^{+}$ amounts to a
maximum of $\sigma_{CD} \approx 450$ mb, to be
compared to an experimental datum of 
$\sigma_{CD}^{exp} = 1.19 \pm 0.11 $ b.

In turn, the experimental data are consistent 
with $J^{\pi}=1/2^{+}$. This can be seen in the
upper part of the figure. There, the result
obtained with the dominant configuration
$\vert ^{18}$C(0$^{+}$)$\otimes 2s_{1/2} \rangle$
and a spectroscopic factor $S_{b} = 0.12$
yields a spectrum compatible with the experimental
data. However, the position of the peak
is located at larger energy in the
experimental spectrum and the width of
the distribution is larger than that resulting
from the calculation. A nice reproduction
of the data can be obtained using the
same configuration with a neutron binding
energy of 0.530 MeV and a spectroscopic
strength $S_{b} = 0.67$, compatible with the
shell model results. This value
of the neutron binding, though larger
than that derived from the present 
mass evaluations ($0.240\pm0.100$ MeV), 
is not at all anomalous. 
In fact, already in the
nuclear breakup experiment of Bazin {\it et al.}
a value of 0.600 MeV was proposed to
reproduce the width of their fragments momentum
distribution data. Therefore, the result 
of the present analysis can be considered 
the confirmation of a higher value 
of the neutron binding energy of $^{19}{\rm C}$.  

\section{Conclusions}
In this work we have shown how the Coulomb
dissociation process furnishes a unique
spectroscopic tool for the investigation of the halo
structure properties of loosely bound nuclear
systems. Nuclear structure model calculations,
such as conventional shell-model approaches
are the necessary guide for the interpretation
of experimental data. Quantitative answers
to some of the questions posed by the
exotic systems encountered in 
far-from-stability lands can be obtained
by a close intercomparison between advanced
experimental utilities with tested
theoretical methodologies.


\section*{Acknowledgments}
Most of the results presented in this work have been
obtained in the frame of a collaboration between
ENEA, RIKEN and The University of Tokyo. I would
like to thank the persons involved in this work:
T.~Tagami, N.~Fukuda, T.~Nakamura T.~Otsuka, and M.~Ishihara.  
I am thankful for the financial support obtained from JSPS 
(Japanese Society for the Promotion of Science) 
and from STA (Science and Technology Agency
of Japan).

\end{document}